# Attack-Graph Threat Modeling Assessment of Ambulatory Medical Devices

Patrick Luckett
University of South Alabama
phl801@jagmail.southalabama.edu

J. Todd McDonald
University of South Alabama
jtmcdonald@southalabama.edu

William Bradley Glisson
University of South Alabama
bglisson@southalabama.edu

**Abstract**

*The continued integration of technology into all aspects of society stresses the need to identify and understand the risk associated with assimilating new technologies. This necessity is heightened when technology is used for medical purposes like ambulatory devices that monitor a patient's vital signs. This integration creates environments that are conducive to malicious activities. The potential impact presents new challenges for the medical community.*

*Hence, this research presents attack graph modeling as a viable solution to identifying vulnerabilities, assessing risk, and forming mitigation strategies to defend ambulatory medical devices from attackers. Common and frequent vulnerabilities and attack strategies related to the various aspects of ambulatory devices, including Bluetooth enabled sensors and Android applications are identified in the literature. Based on this analysis, this research presents an attack graph modeling example on a theoretical device that highlights vulnerabilities and mitigation strategies to consider when designing ambulatory devices with similar components.*

## 1. Introduction

The assimilation of technology into medical related devices is continuing to escalate in today's networked environments. This integration is blatantly visible in Ambulatory Medical Devices (AMDs) and Implantable Medical Devices (IMDs). Patients are able to wear AMDs that can monitor Electrocardiogram (EKG) data to detect arrhythmia, monitor blood glucose levels, administer insulin, and wear pulse oximeters that continuously monitors blood oxygen saturation in real time [40, 55, 56]. Not only does this emerging frontier, potentially, improve the safety and well-being of patients; it also provides a continuous source of data for healthcare practitioners to utilize when they are studying associated disorders.

IMDs, such as infusion pumps, dispense controlled volumes of a drug (e.g. insulin or pain medicine) when it is required by the patient. These implantable drug-delivery systems provide a viable method for achieving remedial drug concentrations in order to enhance patient welfare throughout treatment [23]. Another type of implantable medical device is a pacemaker. Pacemakers are placed under the skin near the heart to stimulate heartbeats [2].

The continued integration of technology into medical devices stresses the need to identify and understand the risk associated with assimilating new technologies. Not only do AMDs and IMDs present a physiological risk to the patients who use the device, but it also presents liability risk to practitioners and businesses who are monitoring and interpreting the data produced by these devices [36]. Environmental issues that increase the risks associated with AMDs and IMDs, when compared to traditional medical devices include accessibility and data transmission modes but these devices are accessible by the patient and the general population while they are in use in everyday activities. In other words, there is no physical tampering restriction imposed by the medical provider, like hospital staff, when these devices are used.

From a data transmission perspective, most communication to and from the device is achieved via a wireless connection by a practitioner who may or may not be in the same location as the device. The type of transmission will vary depending on the solution implemented by the device manufacturer. Some ambulatory devices require a period of data storage, followed by a data upload, while other devices feed a constant stream of data to a storage device while it is in use [44, 50, 51]. These characteristics present opportunities to attackers that are not present in traditional medical devices. Therefore, ambulatory devices should be assessed and modeled independently of the traditional devices and traditional risk models.

From a risk perspective, many risk models have been proposed, investigated and implemented into the health care industry. A few of the traditional models

HICSS

that are commonly discussed include: Failure Mode and Effect Analysis (FMEA) [4], A Risk Management Capability Model for Use in Medical Device Companies [46], and CORAS [43]. However, these models fail to provide concise insight into AMD susceptibility.

The reality is that coupling environmental variable with multiple impact targets creates environments for AMDs and IMDs that entice plausible malicious activities in the areas of data exfiltration, data manipulation, and/or device operation modifications. Hence, this research focuses on adversaries who intentionally attempt to gain unauthorized access to a device for nefarious reasons. In doing so, this research investigates the implementation and use of attack graphs as a viable vehicle for investigating this risk associated with AMDs.

Attack graphs are representations that provide a means of analyzing the susceptibility of a system. These graphs present vulnerabilities, exploits, and conditions for multiple attacks in a single consolidated model that allows for a quantitative examination of each individual attack [7]. A benefit of a graph based model is that it presents a rich view of how vulnerabilities relate to each other.

This paper is organized as follows: Section two investigates the current use of ambulatory devices, as well as their vulnerabilities, risk models, and mitigation strategies. The review of the literature also examines the state of the art in attack graphs and graph modeling. Section three discusses the data sets used for the analysis and section four presents the construction of attack graphs and identification of mitigation strategies. Section five elicits conclusions from the analysis and presents future work.

## 2. Relevant Work

The continued integration of technology into the medical arena has fueled research interest in industry and academia. As this proliferation continues, it can be reasoned that the amount of risk increases due to an increasing attack surface and the introduction of new technology. Recent research indicates that residual data extracted from mobile devices is having an increasing impact in legal environments [3, 16]. The escalating amalgamation of ambulatory medical devices into the healthcare industry forces a need to understand the risk that these devices present to organizations.

### 2.1. Attack Graph Models

There are a number of different styles of attack graphs. A very popular attack graph is the attack tree.

In general, attack trees are directed and acyclic graphs. They express how a specific sequence of attack steps can lead to a system breach. The root node of an attack tree represents the goal of the attacker, and the branches in the tree show the different paths to achieve the goal. The steps to achieve the attack are represented by leaves [2]. Once the graph is built, the probability of achieving an attack can be assigned to nodes or links, and the overall probability of reaching the goal can be found. Attack trees can assess risk to static probabilistic models, time dependent dynamic models, or both [2]. Using the assigned probabilities, the paths with the highest expectation of success can be identified and mitigation strategies can be considered.

Attack trees have been used in a variety of fields to represent security risk and vulnerabilities. The term attack tree was first popularized by Bruce Schneier [53]. They are graphs such that nodes depict attacks and links depict the steps to the goal. The root node is considered to be the goal of the attacker and children of the root are steps needed to achieve this goal. The leafs of the attack tree represent attacks that can no longer be cultivated [48]. Notable application of graph-based attack models include security analysis of supervisory control and data acquisition (SCADA) systems, voting systems, vehicular communication systems, Internet related attacks, and secure software engineering [9].

Alhomidi and Reed [7] used attack graph modeling combined with genetic algorithms to identify the most important security threats on a network. Chen [42] presented a value driven approach to threat modeling based on attack path analysis by introducing stakeholder incentives into commercial off-the-shelf, product vulnerability prioritization.

Kotenko and Chechulin [15] note the major drawback of large attack graphs is computational complexity, and described attack modeling and impact assessment solutions focused on development of attack graph construction and analysis for systems operating in near real-time. Phillips and Swiler [54] state a network-vulnerability risk identification system should be capable of modeling the dynamic conditions of a network. These conditions include the ability of the attacker, concurrent events or attacks, user access controls, and the sequences of attacks that depend on time. Their method uses graph algorithms such as shortest-path to recognize the attack paths with the highest risk.

Louthan et al. [10] describe an approach to modeling hybrid systems, such as programmed control systems and cyber physical systems, that interact with the physical world. Their method used what they term a hybrid attack graph. The hybrid attack graph shows a combined prospective of the space between information systems and a restricted but useful set of hybrid systems that are at risk. Florian et al. [2] state the assets and amount of time available to the attacker and the stepwise execution of complementary attack steps are the central aspects for an attacker in a sophisticated attack. Based on these observations, their paper extends dynamic attack tree models using the ordered parallel behavior of AND-and OR-gates. Vigo et al. [14] proposed an automated attack tree generator using a static analysis approach. The attack trees are automatically inferred from a process of algebraic specification and Satisfiability Modulo Theories in a syntax-directed fashion while avoiding exponential explosion. Their case study used the standard propositional denotation of an attack tree to phrase quantitative problems.

Piètre-Cambacédès et al. [33] note that attack trees are intrinsically static and limited to events that occur independently of each other. They suggest a similar structure based on Boolean logic Driven Markov Processes. This is similar to attack tree models but avoid combinatorial explosions. Roy et al. [34] presented a novel attack tree they refer to as attack countermeasure trees. In their model, defense measures can be posed not only at the leaves of a tree, but any node of the tree. Kordy et al. [31] demonstrated the similarities between attack trees and game theory. They showed attack–defense trees and binary zero-sum two-player extensive form games have proportionate expressive power such that they can be transformed into one another and still preserve the result and architecture.

Attack-defense trees are extensions of attack trees. An attack-defense tree has the same attributes as an attack tree, but also contains defense strategies. Nodes are given characteristics, such as probability, impact, and penalty. This is done in order to enhance the expressive capability of the model. The values of the characteristics are determined based on cognitive assessment and historical events [19].

Kordy et al. [22] compared the computational complexity of attack trees versus attack-defense trees. They identified rules for which extending attack trees did not increase computational complexity. Bagnato et al. [19] also used attack-defense trees, which focus on how attackers and defenders relate, to identify risk to an RFID system in a case study. Based on their model, they were able to identify guidelines to adhere to when using similar strategies.

## 2.2 Risks for Medical Devices

There has also been considerable research in the risk associated with medical devices posed by attackers. Among such devices are implantable medical devices. These devices have become increasingly popular and many are equipped with wireless communications which make them prime targets for attackers [32]. In the article, *Researchers fight to keep implanted medical devices safe from hackers,* Leavitt [32] notes that over two million people in the US have an implantable medical device. Many of these devices communicate using wireless capability. Also noted in the article were the researchers from Harvard University, University of Massachusetts Amherst, and University of Washington who were able to hijack the short-range signals that an implantable cardiac defibrillator sent to a legitimate independent controller and caused it to emit a shock capable of inducing a fatal heart rhythm [32].

Arney et al. [26] state that adversaries who attack medical devices can be classified into two categories, active and passive. Active adversaries have the ability to spy on communications among devices, network controllers and supervisors. They are then able to insert messages, spoof, and damage the integrity of the device. The second type of adversaries, passive, eavesdrop for the purposes of acquiring private data stored in a device. They also note four classes of targets that adversaries attack within medical device systems: patient physical security, patient data security (privacy), medical device physical security, and data security of the health-care institution that deploys the device [26].

**Table 1. IEEE 802.15.6 Communication [38]**

| Implant to Implant | 402-405 MHz |
|---|---|
| Implant to Body Surface | 402-405 MHz |
| Implant to External | 402-405 MHz |
| Body Surface to Body Surface (LOS) | 13.5, 50, 400, 600, 900 MHz, 2.4, 3.1 - 10.6 GHz |
| Body Surface to Body Surface (NLOS) | 13.5, 50, 400, 600, 900 MHz, 2.4, 3.1 - 10.6 GHz |
| Body Surface to External (LOS) | 13.5, 50, 400, 600, 900 MHz, 2.4, 3.1 - 10.6 GHz |
| Body Surface to External (NLOS) | 13.5, 50, 400, 600, 900 MHz, 2.4, 3.1 - 10.6 GHz |

Burleson et al. [20] note that threat modeling is vital to assessing the security vulnerabilities to medical devices, and the risk posed by the vulnerabilities varies along with the nature of the data or the ramification of actuation. Radcliffe [39] was able to reverse engineer an insulin pump's packet structure. His research showed the insulin pump did not encrypt the medical data it transmitted and did not authenticate the components that were communicating. Li [27] was able to take control of an insulin pump, including the ability to terminate transmission of insulin or inject large amounts of insulin, and suggested mitigation strategies using rolling-code cryptographic protocols and body-coupled communication.

Xu [5] created an automated attack trees generator for implantable medical devices using process modeling and hazard analysis. He also demonstrated its use on Patient Controlled Analgesia, which is used for delivering pain medication to patients in hospitals. Rushanan and Kune [12]note the security of the telemetry interface on implantable medical devices has received much attention in the academic community, but the risk of software exploitation and the sensor interface layer requires further research. Rostami et al. [18] describe the challenge in securing medical devices, including inability to use common approaches such as passwords and certificates because practitioners would not have access to the device in an emergency setting, and implantable medical devices are limited in power consumption and computational capability, which limits security strategies. As stated, the intention of their paper was to stimulate further research in the areas of implantable medical device security and medical-device security in general.

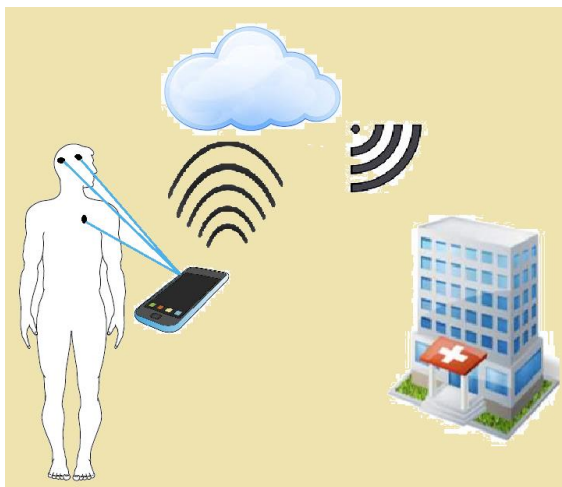

**Figure 1. Hypothetical Network**

### 2.3 Wireless Body Area Network

Ambulatory devices can be a single unit that may or may not transmit data, or they can be one of many devices that make up a wireless body area network (WBAN). Body area networks are localized wireless networks that have the ability to support a wide variety of medical devices [28]. A wireless body area network can consist of devices to monitor physiological data, devices to display collected data, devices to perform calculations, devices to administer medication, and devices to store the collected data. IEEE 802.15 is concerned with the development of agreeable standards for Personal Area Networks or short distance wireless networks. It addresses wireless networking of ambulatory computing devices such as PCs, cell phones, and consumer electronics [18]. The IEEE standard 802.15.6 is the latest standard for wireless body area networks. This standard specifies short range wireless communication inside or on the human body [28]. However, several security problems have been noted [4], and communication can be achieved in a variety of ways, including ZigBee, Bluetooth, internet, WIMAX, RF, Volte, and 2, 3, or 4G mobile telephone networks [30, 47, 49]. Wireless body area networks function in either a one-hop or two-hop star topology [35]. Table 1 – IEEE 802.15.6 Communication describes the various communication channels laid out by IEEE 802.15.6.

The IEEE 802.15.6 standard identifies a security paradigm for wireless body area networks that defines three levels of security [37]:
1. Unsecured Communication - Data transmitted in unsecured frames. Provides no measure for integrity, validation, authenticity, replay defense, privacy, and confidentiality.
2. Authentication/ no Encryption- Data that is transmitted is authenticated but not encrypted.
3. Authentication/ Encryption- Data transmitted is authenticated and encrypted.

All devices in a wireless body area network fall into one of these three categories.

A significant amount of research has been conducted on attack graphs and risks to medical devices and body area networks. This research suggests combining the two to assess risk to ambulatory medical devices and form mitigation strategies.

### 3. Data and Model

In order to display the use of attack graphs and form mitigation strategies, this research uses a model of a theoretical ambulatory device as seen in Figure 1 – Hypothetical Wireless body Network. The model is referred to as theoretical because it is currently in production and, therefore, not yet available for proper testing. The device is a wireless body network that consist of three sensors. Two sensors are worn on the head and one sensor is worn on the chest. These types of sensors are commercially available, and capable of monitoring various biological data, including heart rate, EEG signals, or body temperature. The sensors communicate through a wireless signal to a cellular smart phone which runs an application that processes, analyzes, and stores the data.

**Table 2. Bluetooth Attacks and Mitigation**

| Author | Attack | Mitigation Strategy |
|---|---|---|
| Padgette / Minar [41] [24] | Capture Bluetooth device address | Set device to lowest power level |
| Minar [24] | BluePrinting | Keep device address secret |
| Minar [24] | Reflection attack | Use encryption, Keep device address secret |
| Padgette [41] | Repeatable authentication attempts | Limit authentication request, Set device to lowest power |
| Minar [24] | Blueover | Keep device address secret |
| Padgette [41] | Static SSP pass keys | Random, passkeys at each pairing |
| Padgette / Minar [41] [24] | Encryption key negotiable. | Full 128 bit key, establish min key size |
| Padgette / Dardanelli [41] | No authentication | Application level security |
| Minar / Panse [24] [25] | Bluesnarfing | non-discover mode |
| Minar / Panse [24] [25] | Pin Cracking | Use random long pin codes |
| Minar [24] | MIM/Impersonation Attack | Link encryption, Link keys based on combination keys, Security mode 3, Set device to lowest power |
| Minar [24] | Pairing Eavesdropping | Pair as little as possible, Link encryption, Set device to lowest power |

The application allocates a specific amount of memory for data storage and uploads the data to cloud storage when needed. Doctors have access to the cloud storage for data analysis. The application analyzes the data and, if an anomaly occurs, it sends a text message to the patient and patient's emergency contact as well as an email to the patient's doctor. Due to industry popularity [1], the scope of this research focuses on a smartphone running an Android Operating System that utilizes Bluetooth to communicate with the sensors. Bluetooth is a short range (10-100m) low power wireless technology that operates from 2.4 to 2.4835 GHz at a data rate of 1, 2, or 3 Mbps [6]. Three basic security services provided in the Bluetooth standard and identified in a National Institute of Standards and Technology (NIST) guide include authentication, confidentiality and authorization [41]. The report defines authentication as the ability to identify communicating devices through a unique device address. The report notes that the Bluetooth standard does not support user authentication natively. Confidentiality focuses on averting the compromise of information by ensuring that only authorized devices have access to transmitted data. Authorization concentrates on resource governance based on device authorization prior to sanctioning interaction.

In general, Bluetooth security threats can be grouped into three categories that include disclosure, integrity and denial. A disclosure threat occurs when information is leaked from the system, an integrity threat is when an attacker deliberately alters data to fool the receiver and a denial of service threat occurs when an attacker is able to limit a user's access to a device or application [57]. This research focuses on data acquisition and/or manipulation. A literature review was performed to identify appropriate attack strategies on individual vectors in the theoretical model. Table 2 – Bluetooth Attacks and Mitigation summarizes attacks and provides mitigation strategies when available on Bluetooth enabled devices, and Table 3 – Android Attacks and Mitigation provides attacks and mitigation strategies on devices running an Android operating system. In both tables the first column lists the author and reference. The second column gives the title or style of an attack and the last column provides a of list possible mitigation strategies.

This research assumes attackers are capable enough to acquire information regarding communication frequency and modulation. This information is, generally, easily found in an online copy of a device's user manual or by searching for the specific device on the Federal Communication Commission website. Therefore, the reconnaissance steps are omitted in the model.

**Table 3. Android Attacks and Mitigation**

| Author | Attack | Mitigation Strategy |
|---|---|---|
| Vidas [29] | Physical Attack | User Authentication |
| Vidas / Enck [29] [21] | Permission Model Attack | App certified |
| Chen [8] | UI State Inference Attacks | File System Access Control, Buffer Reuse |
| Noor [17] | Man In The Middle | Encryption, No default password |
| Oli [11] | General | No automatic connection to Wi-Fi, Disable Wi-Fi when not in use |
| Dondyk [65] | Denial of Service | Disable Wi-Fi when not in use |

## 4. Attack Graphs

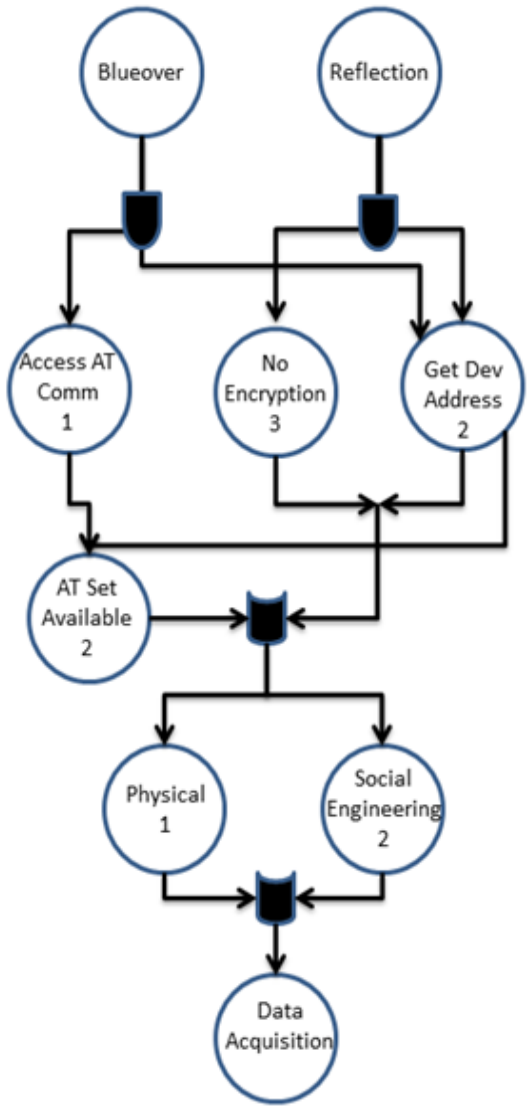

**Figure 2. Attack Graph**

An inherent difficulty with attack graph modeling is assigning weights to the edges of the graph [45]. Specifically, what method does one us to assign a numerical value to an attack that cannot be quantitatively assessed? This research presents an alternative approach to assigning weights to links in the graph. The graph, as seen in Figure 2 - Bluetooth Attack Graph, assigns a numerical value to nodes. This value represents the likelihood of achieving success in a given node. By adding the value of each node in the path and dividing by the number of nodes traversed in the path, an 'average' risk is assigned for the attack. Node risks are assigned based on the following concepts:

1. Monotonicity as stated by Amman et al "means that no action an attacker takes interferes with the attacker's ability to take any other action" [52]. Hence, any calculations derived from the attacks must consider all attack vectors.
2. The frequency concept simply means that increased recurrence is displayed via increased node weight [45]. Nodes that are visited more often are given higher risk. This is not because it is easily achieved, but because it is a vital step to many different attacks.
3. Complexity refers to the difficulty of an attack. For example, BlueSnarfing is described in literature as "the software tools required to steal information from Bluetooth enabled mobile phones (that) are widely available in the Web" [24], therefore an "equipment" node would be assigned a higher weight because it is easily achieved.

Figure 2 is an example attack graph on the theoretical device. Here, the goal of the attacker is data acquisition. Blueover is an attack used to acquire sensitive or private data from a mobile device equipped with Bluetooth. Reflection attacks are a type of 'man in the middle attack' against Bluetooth enabled devices. 'Access AT Comm' refers to an attacker having access to the address translation command. 'Get Dev Address' refers to the ability to get the Bluetooth device address, and 'No Encryption' means the communication between devices is not encrypted. 'AT Set Available' means the mode of address translation command is set to available. Finally, physical refers to an attack where an attacker gets physical access to a device, and social engineering is when the attacker uses methods such as phishing to get the needed information. Attacks were selected from those in tables 2 and 3.

The attack graph depicted in Figure 2 highlights two types of attacks, Blueover and reflection. Blueover requires two initial steps for success. The first is 'Get Device Address', and the second is 'Access AT Comm'. The symbol on the graph connecting the two links is an 'And' symbol, which means both must be achieved. The next step for 'Access AT Comm' is 'AT set Available', which means the mode of the address translation must be set too available.

The next step requires either a physical attack or social engineering to achieve the goal of data acquisition. The two initial steps for a reflection attack are 'Get Device Address' and 'No Encryption'. Once this has taken place, either a physical attack or a social engineering attack can be instigated.

By modeling attacks in this manner, it is easy to identify the most important security issues. For example, most paths eventually go through social engineering or physical nodes as depicted in Figure 2.

For the purposes of this discussion, the attack nodes illustrated in Figure 2 are considered a moderate risk. While the US Food and Drug Administration (FDA) suggests actions that manufacturers should consider in order to secure medical devices such as "Limiting access to devices to trusted users through the use of authentication, such as ID and password, smart card and biometrics, including multi-layered authentication" [13], the guidance is only a recommendation and does not establish legally enforceable responsibilities.

It is reasonable to assume administrators will attempt to hold medical devices, which reside in a hospital setting, to the guidelines that are set forth by the FDA. However, such an assumption should not be made for third party software that runs on devices such as mobile phones.

Another consideration is that ambulatory devices could have suggested authentication protocols; however, there is no guarantee that users who do not understand the possible risk will enable authentication, or use reasonable passwords to protect devices. Since these risk are considered moderate, it would be a good idea to educate patients on the dangers associated with these types of attacks along with how to avoid them. Another node frequently visited is 'Get Dev Add', which stands for Get the Bluetooth Device Address. Hence, it would be a good idea for patients to ensure that the Bluetooth Device Address stays concealed.

Table 4 - Blueover Possible Paths presents a list of the possible attack paths for a Blueover attack. The table only depicts attacks that could lead to success. For example, both 'Access AT Command' and 'Get Device Address' must be achieved. In any case where one of these attacks fails, the probability of success is zero. Therefore, those attacks are not listed. This indicates that mitigation strategies relating to those nodes should be top priority. An analysis of the table data indicates that the probability of the success of an attack is reduced or increased by removing the threat to any individual node corresponding to an 'or' gate.

The logic for Tables 4 and 5 are derived via the following calculation. Each row in the table has an S followed by a number. The S stands for success while the number is the assigned weight from the node. If all of the attacks are successful, the weighted impact totals eight. If an attack is not achieved, the S is turned into an F (for failure) and the weight is assigned a zero to reduce the likelihood of achieving the overall attack goal. The total possible value in both tables is eight. This value represents the value of treating every node as a success and summing the values. The probabilities between the tables appear to correspond, but table 5 has one less intermittent step ('AT Set Available').

**Table 4. Blueover Possible Paths**

| Access AT Comm | Get Dev Add | AT Set Aval | P | Soc | Norm | Goal |
|---|---|---|---|---|---|---|
| S-1 | S-2 | S-2 | S-1 | S-2 | 8/8 | 100% |
| S-1 | S-2 | S-2 | S-1 | F-0 | 6/8 | 75% |
| S-1 | S-2 | S-2 | F-0 | S-2 | 7/8 | 87% |
| S-1 | S-2 | F-0 | S-1 | S-2 | 6/8 | 75% |
| S-1 | S-2 | F-0 | S-1 | F-0 | 4/8 | 50% |
| S-1 | S-2 | F-0 | F-0 | S-2 | 5/8 | 63% |

Key: S=Success, F=Failure, P=Physical Attack, Soc= Social Engineering, Norm=Actual/Potential, Goal= Probability of success if given steps are achieved.

Table 5 - Reflection Attack Possible Paths shows the possible attack paths for a Reflection attack. Again, 'No Encryption' and 'Get Device Address' are both required for success, so only paths with attacks that are successful are shown. The evaluation of this table indicates that social engineering attacks should be addressed before physical attacks. This is due to a higher probability of achieving the attack goal is higher for social engineering versus a physical attack.

Viewing both tables together gives further insight into common attack vectors. For example, the node 'Get Device Address' is required in both attacks. 'Get Device Address' refers to the ability to get the Bluetooth device address. Since this attack goal is pursued in two different attack types, all Bluetooth device users should take steps to keep device addresses secret. Evaluations should also take into account the attack paths.

**Table 5. Reflection Attack Possible Paths**

| No Encryption | Get Dev Add. | P | Soc | Norm | Goal |
|---|---|---|---|---|---|
| S-3 | S-2 | S-1 | S-2 | 8/8 | 100% |
| S-3 | S-2 | S-1 | F-0 | 6/8 | 75% |
| S-3 | S-2 | F-0 | S-2 | 7/8 | 87% |

Key: S=Success, F=Failure, P=Physical Attack, Soc= Social Engineering, Norm=Actual/Potential, Goal= Probability of success if given steps are achieved.

Reviewers should consider the number of steps in a path. In most cases one would assume a shorter path is easier to achieve. However, the correspondence between the probabilities on the tables for particular paths is also due to the 'No Encryption' node. The 'No Encryption' node has a very high risk value, which offsets the fact that the attack has fewer steps. In any case, producing attack graphs and the corresponding attack path tables provides detailed insight on the vulnerabilities and possible mitigation strategies within a system.

These models highlight the need to assess risk to ambulatory medical devices independently of traditional medical devices. Vulnerabilities such as physical access and social engineering would have less probability of success for traditional medical devices for various reasons. Traditional devices in a hospital setting are generally monitored by the hospital staff, making the success of a physical attack less likely. Hospital personnel receive training on the use and maintenance of medical devices, making them less likely to fall victim to a social engineering attack. Devices in hospital settings are generally 'hard wired' or they are on a private network. In addition, many devices use proprietary software. This makes threats such as 'Access AT Comm' less likely, if not impossible.

## 5. Conclusion and Future Work

Ambulatory medical devices offer a viable alternative for patients who require constant monitoring. These devices provide a means for administering medication, monitoring vital signs, and improving a patient's overall quality of life. However, as with any technology, it is important to understand the risk associated with the use of these devices. This is especially important for ambulatory medical devices, which can have direct or indirect impact on a patient's health and wellbeing.

Attack graphs offer a visual approach to identifying risk within complex systems. The steps required to achieve an attack are easily identifiable using this approach. Hence, the identification of attacks aids designers in developing mitigation strategies to prevent the successful execution of an attack.

This research demonstrates attack graph modeling on a theoretical ambulatory medical device. The theoretical device contains components and software that is common among ambulatory devices today. This research highlights the need to model ambulatory devices separately from traditional medical devices by demonstrating certain attack vectors that pose greater risk to ambulatory devices, such as physical attacks and social engineering. To our knowledge, this is the first time attack graph modeling has been used for ambulatory medical devices.

Additional future work will consider the architecture of the attack graph. In this research, weights were assigned to the nodes of the graph. Future work will examine the impact of assigning weights to the links between nodes along with developing combined weighting systems in order to identify which style of attack graph is the most appropriate for ambulatory medical devices. Once modeling is complete, mitigation strategies will be identified and tested.

## 6. References

[1] Market Share Held by Smartphone Platforms in the United States, http://www.statista.com 2016.

[2] Arnold, Florian, Dennis Guck, Rajesh Kumar, and Mariële Stoelinga, "Sequential and Parallel Attack Tree Modelling": Computer Safety, Reliability, and Security, Springer, 2015, pp. 291-299.

[3] Berman, Kiyoshi, William Bradley Glisson, and L. Milton Glisson, "Investigating the Impact of Global Positioning System (Gps) Evidence in Court Cases", Hawaii International Conference on System Sciences (HICSS-48), 2015

[4] Toorani, Mohsen, "On Vulnerabilities of the Security Association in the Ieee 802.15. 6 Standard": Financial Cryptography and Data Security, Springer, 2015, pp. 245-260.

[5] Xu, Jian, Systematic Vulnerability Evaluation of Interoperable Medical Device System Using Attack Trees, WORCESTER POLYTECHNIC INSTITUTE, 2015.

[6] Practical Testing of Bluetooth Devices, http://www.litepoint.com/wp-content/uploads/2014/02/ Practical-Testing-of-Bluetooth-Devices_WhitePaper.pdf

[7] Alhomidi, Mohammed, and Martin Reed, "Attack Graph-Based Risk Assessment and Optimisation Approach", International Journal of Network Security & Its Applications, 6(3), 2014, pp. 31.

[8] Chen, Qi Alfred, Zhiyun Qian, and Z Morley Mao, "Peeking into Your App without Actually Seeing It: Ui State Inference and Novel Android Attacks", 2014, pp. 1037-1052.

[9] Kordy, Barbara, Ludovic Piètre-Cambacédès, and Patrick Schweitzer, "Dag-Based Attack and Defense Modeling: Don't Miss the Forest for the Attack Trees", Computer science review, 13(2014, pp. 1-38.

[10] Louthan, George, Michael Haney, Phoebe Hardwicke, Peter Hawrylak, and John Hale, "Hybrid Extensions for Stateful Attack Graphs", ACM, 2014, pp. 101-104.

[11] Oli, Vckp, and Elayaraja Ponram, "Wireless Fidelity Real Time Security System", arXiv preprint arXiv:1405.1019, 2014,

[12] Rushanan, Michael, Aviel D Rubin, Denis Foo Kune, and Colleen M Swanson, "Sok: Security and Privacy in Implantable Medical Devices and Body Area Networks", IEEE, 2014, pp. 524-539.

[13] Us Food and Drug Administration, "Content of Premarket Submissions for Management of Cybersecurity in Medical Devices: Draft Guidance for Industry and Food and Drug Administration Staff", 2014, pp. 9.

[14] Vigo, Roberto, Flemming Nielson, and Hanne Riis Nielson, "Automated Generation of Attack Trees", IEEE, 2014, pp. 337-350.

[15] Kotenko, Igor, and Andrey Chechulin, "A Cyber Attack Modeling and Impact Assessment Framework", IEEE, 2013, pp. 1-24.

[16] Mcmillan, Jack, William Bradley Glisson, and Michael Bromby, "Investigating the Increase in Mobile Phone Evidence in Criminal Activities", Hawaii International Conference on System Sciences (HICSS-46), 2013

[17] Noor, Mardiana Mohamad, and Wan Haslina Hassan, "Wireless Networks: Developments, Threats and Countermeasures", International Journal of Digital Information and Wireless Communications (IJDIWC), 3(1), 2013, pp. 125-140.

[18] Rostami, Masoud, Wayne Burleson, Farinaz Koushanfar, and Ari Juels, "Balancing Security and Utility in Medical Devices?", ACM, 2013, pp. 13.

[19] Bagnato, Alessandra, Barbara Kordy, Per Håkon Meland, and Patrick Schweitzer, "Attribute Decoration of Attack–Defense Trees", International Journal of Secure Software Engineering (IJSSE), 3(2), 2012, pp. 1-35.

[20] Burleson, Wayne, Shane S Clark, Benjamin Ransford, and Kevin Fu, "Design Challenges for Secure Implantable Medical Devices", ACM, 2012, pp. 12-17.

[21] Crussell, Jonathan, Clint Gibler, and Hao Chen, "Attack of the Clones: Detecting Cloned Applications on Android Markets": Computer Security–Esorics 2012, Springer, 2012, pp. 37-54.

[22] Kordy, Barbara, Marc Pouly, and Patrick Schweitzer, "Computational Aspects of Attack–Defense Trees": Security and Intelligent Information Systems, Springer, 2012, pp. 103-116.

[23] Meng, Ellis, and Tuan Hoang, "Micro-and Nano-Fabricated Implantable Drug-Delivery Systems", Therapeutic delivery, 3(12), 2012, pp. 1457-1467.

[24] Minar, Nateq Be-Nazir Ibn, and Mohammed Tarique, "Bluetooth Security Threats and Solutions: A Survey", International Journal of Distributed and Parallel Systems, 3(1), 2012, pp. 127.

[25] Panse, Trishna, and Vivek Kapoor, "A Review on Security Mechanism of Bluetooth Communication", International Journal of Computer Science and Information Technologies, 3(2), 2012, pp. 3419-3422.

[26] Arney, David, Krishna K Venkatasubramanian, Oleg Sokolsky, and Insup Lee, "Biomedical Devices and Systems Security", IEEE, 2011, pp. 2376-2379.

[27] Li, Chunxiao, Anand Raghunathan, and Niraj K Jha, "Hijacking an Insulin Pump: Security Attacks and Defenses for a Diabetes Therapy System", IEEE, 2011, pp. 150-156.

[28] Body Area Networks: A Way to Improve Remote Patient Monitering, www.ecnmag.com/article/2011/11/body-area-networks-way-improve-remote-patient-monitoring 2011.

[29] Vidas, Timothy, Daniel Votipka, and Nicolas Christin, "All Your Droid Are Belong to Us: A Survey of Current Android Attacks", 2011, pp. 81-90.

[30] Fereydouni-Forouzandeh, Fariborz, Ultra Low Energy Communication Protocol for Implantable Wireless Body Sensor Networks, Concordia University, 2010.

[31] Kordy, Barbara, Sjouke Mauw, Matthijs Melissen, and Patrick Schweitzer, "Attack–Defense Trees and Two-Player Binary Zero-Sum Extensive Form Games Are Equivalent": Decision and Game Theory for Security, Springer, 2010, pp. 245-256.

[32] Leavitt, Neal, "Researchers Fight to Keep Implanted Medical Devices Safe from Hackers", Computer8), 2010, pp. 11-14.

[33] Piètre-Cambacédès, Ludovic, and Marc Bouissou, "Beyond Attack Trees: Dynamic Security Modeling with Boolean Logic Driven Markov Processes (Bdmp)", IEEE, 2010, pp. 199-208.

[34] Roy, Arpan, Dong Seong Kim, and Kishor S Trivedi, "Cyber Security Analysis Using Attack Countermeasure Trees", ACM, 2010, pp. 28.

[35] Tachtatzis, Christos, Fabio Di Franco, David C Tracey, Nick F Timmons, and Jim Morrison, "An Energy Analysis of Ieee 802.15. 6 Scheduled Access Modes", IEEE, 2010, pp. 1270-1275.

[36] Zimetbaum, Peter, and Alena Goldman, "Ambulatory Arrhythmia Monitoring Choosing the Right Device", Circulation, 122(16), 2010, pp. 1629-1636.

[37] "Ieee P802.15 Working Group for Wireless Personal Area Networks (Wpans): Medwin Mac and Security Proposal Documentation", IEEE, IEEE, 2009

[38] Yazdandoost, K, and K Sayrafian, "Channel Model for Body Area Network (Ban), Ieee P802. 15-08-0780-09-0006", IEEE 802.15 Working Group Document, 2009,

[39] Halperin, Daniel, Thomas S Heydt-Benjamin, Benjamin Ransford, Shane S Clark, Benessa Defend, Will Morgan, Kevin Fu, Tadayoshi Kohno, and William H Maisel, "Pacemakers and Implantable Cardiac Defibrillators: Software Radio Attacks and Zero-Power Defenses", IEEE, 2008, pp. 129-142.

[40] Lee, Young-Dong, Sang-Joong Jung, Yong-Su Seo, and Wan-Young Chung, "Measurement of Motion Activity During Ambulatory Using Pulse Oximeter and Triaxial Accelerometer", IEEE, 2008, pp. 436-441.

[41] Scarfone, Karen, and John Padgette, "Guide to Bluetooth Security", NIST Special Publication, 800(2008, pp. 121.

[42] Chen, Yue, Software Security Economics and Threat Modeling Based on Attack Path Analysis; a Stakeholder Value Driven Approach, ProQuest, 2007.

[43] Den Braber, Folker, Ida Hogganvik, Ms Lund, Ketik Stølen, and Fredrik Vraalsen, "Model-Based Security Analysis in Seven Steps—a Guided Tour to the Coras Method", BT Technology Journal, 25(1), 2007, pp. 101-117.

[44] Kumar, Kishore, and Laxmikant Rashinkar, "System and Method for Real-Time Monitoring, Assessment, Analysis, Retrieval, and Storage of Physiological Data over a Wide Area Network", Google Patents, 2007

[45] Wang, Lingyu, Anoop Singhal, and Sushil Jajodia, "Toward Measuring Network Security Using Attack Graphs", ACM, 2007, pp. 49-54.

[46] Burton, John, Fergal Mccaffery, and Ita Richardson, "A Risk Management Capability Model for Use in Medical Device Companies", ACM, 2006, pp. 3-8.

[47] Otto, Chris, Aleksandar Milenkovic, Corey Sanders, and Emil Jovanov, "System Architecture of a Wireless Body Area Sensor Network for Ubiquitous Health Monitoring", Journal of mobile multimedia, 1(4), 2006, pp. 307-326.

[48] Mauw, Sjouke, and Martijn Oostdijk, "Foundations of Attack Trees": Information Security and Cryptology-Icisc 2005, Springer, 2005, pp. 186-198.

[49] Baldus, Heribert, Karin Klabunde, and Guido Muesch, "Reliable Set-up of Medical Body-Sensor Networks": Wireless Sensor Networks, Springer, 2004, pp. 353-363.

[50] Lang, Brook W, "Emergency Medical Treatment System", Google Patents, 2004

[51] Linder, Marshal, and Thomas Kaib, "Data Collection and System Management for Patient-Worn Medical Devices", Google Patents, 2004

[52] Ammann, Paul, Duminda Wijesekera, and Saket Kaushik, "Scalable, Graph-Based Network Vulnerability Analysis", Proceedings of the 9th ACM conference on Computer and communications security, 2002, pp. 217-224.

[53] Schneier, Bruce, "Attack Trees", Dr. Dobb's journal, 24(12), 1999, pp. 21-29.

[54] Phillips, Cynthia, and Laura Painton Swiler, "A Graph-Based System for Network-Vulnerability Analysis", ACM, 1998, pp. 71-79.

[55] Araki, Haruo, Y Koiwaya, O Nakagaki, and M Nakamura, "Diurnal Distribution of St-Segment Elevation and Related Arrhythmias in Patients with Variant Angina: A Study by Ambulatory Ecg Monitoring", Circulation, 67(5), 1983, pp. 995-1000.

[56] Rupp, William M, Jose J Barbosa, Perry J Blackshear, Hildreth B Mccarthy, Thomas D Rohde, Fay J Goldenberg, Thomas G Rublein, Frank D Dorman, and Henry Buchwald, "The Use of an Implantable Insulin Pump in the Treatment of Type Ii Diabetes", New England Journal of Medicine, 307(5), 1982, pp. 265-270.

[57] Laurie, Adam, Marcel Holtmann, and Martin Herfurt, "The Bluebug", AL Digital Ltd. http://trifinite.org/trifinite_stuff_bluebug. html